\documentclass[useAMS,usenatbib]{mn2e}
\usepackage{graphicx,amssym}
\newcommand{\bc}{\begin{center}}
\newcommand{\ec}{\end{center}}

\title[Galaxy growth in the concordance $\Lambda$CDM cosmology]
      {Galaxy growth in the concordance $\Lambda$CDM cosmology}
\author[Qi Guo and Simon D. M. White]
       {Qi Guo$^{1}$\thanks{Email: guoqi@mpa-garching.mpg.de},
	Simon D.M. White$^{1}$
        \\     
        $^1$ Max Planck Institut f\"ur Astrophysik, 
        }
\begin{document}

\date{Accepted  ???? ??. 
      Received  ???? ??; 
      in original form 2006 ???? ??}

\pagerange{\pageref{firstpage}--\pageref{lastpage}} 
\pubyear{200?}

\maketitle

\label{firstpage}

\begin{abstract}
We use galaxy and dark halo data from the public database for the
Millennium Simulation to study the growth of galaxies in the
\citet{Lucia2006} model for galaxy formation. Previous work has shown
this model to reproduce many aspects of the systematic properties and
the clustering of real galaxies, both in the nearby universe and at
high redshift. It assumes the stellar masses of galaxies to increase
through three processes, major mergers, the accretion of smaller
satellite systems, and star formation. We show the relative importance
of these three modes to be a strong function of stellar mass and of
redshift.  Galaxy growth through major mergers depends strongly on
stellar mass, but only weakly on redshift. Except for massive
systems, minor mergers contribute more to galaxy growth than major
mergers at all redshifts and at all stellar masses. For galaxies
significantly less massive than the Milky Way, star formation
dominates the growth at all epochs. For galaxies significantly more
massive than the Milky Way, growth through mergers is the dominant
process at all epochs. At a stellar mass of $6\times 10^{10}M_\odot$,
star formation dominates at $z>1$ and mergers at later times. At every
stellar mass, the growth rates through star formation increase rapidly
with increasing redshift.  Specific star formation rates are a
decreasing function of stellar mass not only at $z=0$ but also at all
higher redshifts. For comparison, we carry out a similar analysis of
the growth of dark matter halos. In contrast to the galaxies, growth
rates depend strongly on redshift, but only weakly on mass. They agree
qualitatively with analytic predictions for halo growth.

\end{abstract}


%%%%%%%%%%%%%%%%%%%%%%%%%%%%%%%%%%%%%%%%%%%%%%%%%%%%%%%%%%%%%%%%%%%%%%%%%%%%%%
\section{Introduction}
\label{sec:intro}

Galaxy mergers play an important role in galaxy formation and
evolution.  They add new gas and stars. They drive gas motions which
feed starbursts and central supermassive black holes, and, for
comparably massive systems, they entirely restructure both
galaxies. \cite{1976T} was the first to stress that the abundance of
tidally distorted spirals in the nearby universe suggests that ``star
piles'' produced by past interactions might account for the majority
of observed elliptical galaxies. \cite{1978W} carried out the first
dynamically consistent 3-dimensional simulations showing that mergers
do indeed produce remnants with a structure similar to that of
ellipticals, a conclusion which has been reinforced by increasingly
realistic simulations of purely stellar systems \citep{FS1982,
Barnes1988, NB2003}. Inclusion of the gas component showed that a
substantial fraction of the interstellar medium should be driven to
the centre in major mergers \citep{NW1983, BH1991, MH1996}. This work
supported the identification of ultraluminous infrared galaxies as
merging systems \citep{Sanders1988} but led to remnant galaxies with
cores which are denser than observed ellipticals.  Recent work
suggests that this contradiction may be resolved by strong AGN- or
starburst-generated winds which expel a large fraction of the gas from
the galaxy \citep{volker2005b,Matteo2005}. Work on mergers of unequal
galaxies suggests that while such mergers may not greatly alter the
structure of the larger system \citep{VW1999, Abadi2003} they can
nevertheless stimulate substantial rearrangements of its gas with
associated star formation and AGN activity \citep{MH1994}.

In the standard $\Lambda$CDM cosmology structure forms
hierarchically. Small dark matter halos form first and then aggregate
into progressively larger systems. At any given time cosmic matter is
distributed over nonlinear objects spanning many decades in mass, and
growth is driven by merging with similar halos, by
accretion of much smaller halos and of diffuse material, and by
destruction by infall onto larger halos \citep[e.g.][]{LC1993}. The
situation is made more complex by the fact that the inner cores of
halos often survive as long-lived substructure within the larger
objects by which they are accreted \citep{GG,Moore1999,
Gao2004}. Galaxies form at the centres of halos in the way suggested
by \cite{WR1978} and are swept along with the growth of dark matter
structure. They gain stars through formation from their interstellar
medium, which may be replenished by infall from their surroundings,
and by incorporating the stars of galaxies which merge with them. The
interaction between these processes drives the overall evolution of
the population and cannot be followed without treating the associated
baryonic astrophysics (gas condensation, formation and evolution of
stars and black holes, feedback from supernovae and AGN, chemical
enrichment, production of observable radiation etc.).

Early studies of the evolution of the galaxy population embedded
simplifed treatments of this baryonic physics in Monte Carlo
realisations of the merger trees associated with the formation of
individual dark halos \citep{KGW1993, Cole1994, Cole2000, SP1999}. The
spatial distribution of galaxies could then be studied using the halo
distribution from an N-body simulation of structure formation
\citep{KNS1997, Benson2000}. Improvements on this scheme have used
higher resolution N-body simulations so that the merging trees can
taken directly from the simulation itself, thereby allowing the
evolution of the galaxy population to be followed in a single
consistent simulation \citep{Kauffmann1999,Springel2001b,
Helly2003,Hatton2003,Nature2005,Kang2005}. A parallel approach has
followed the dynamics of diffuse gas (in particular, aspects of the
gas condensation and galactic wind processes) by adding a hydrodynamic
scheme to the N-body treatment of dark matter while continuing to
treat star formation and evolution by semi-analytic means
\citep{CO1992, NW1994, KWH1996}.  The development path here has
involved continual improvement of the simulation schemes to increase
resolution and to treat the accessible physics more realistically \citep[e.g.][]{CO2000, SH2003, Pfrommer2006}.  Recent work
in both approaches has focussed on how feedback from AGN may regulate
the formation and evolution of their host galaxies \citep{volker2005b,
Matteo2005,Croton2006, Bower2006}.

This body of work has demonstrated that while galaxy mergers are an
important aspect of the evolution of the galaxy population, they do
not simply parallel the mergers of dark halos.  As \cite{WR1978}
stressed, galaxies must remain distinct after the merger of their
halos if we are to understand the formation of galaxy
clusters. \cite{Fall1979} noted that late-type giant galaxies cannot
have undergone recent major mergers since these would destroy their
stellar disks. While many more recent studies have followed
\cite{1976T} in arguing that massive elliptical galaxies assembled
relatively recently through mergers  \citep[e.g.][]{KC1998, Dokkum2005,
Lucia2006} other authors have used the age and uniformity of their
stellar populations and their apparently undiminished abundance at
high redshift to argue against such late assembly
\citep[e.g.][]{CDR2006}. Observational estimates of merging rates, based
primarily on counts of very close pairs of galaxies, or of
morphological evidence for recent merging, have varied widely due to
uncertainties in the associated timescales \citep{Le2000,Lin2004}. In
addition, attempts to measure the evolution of the merger rate,
usually parametrised as proportional to $(1+z)^\alpha$ have obtained
values for the exponent $\alpha$ ranging from 0 to
6. \citep{Bell2006,Carlberg2000,Patton2002,Conselice2003,Bundy2004,Lin2004}.

In the present paper we analyse the build-up of the galaxy population
in the galaxy formation model of \cite{LB2007} which is implemented on
the very large Millennium Simulation \citep{Nature2005}. This model
updates that of \citep{Croton2006} and is a reasonable match to the
clustering and to many of the systematic properties of the local
galaxy population. It is also consistent with most available data at
high-redshift \citep{KW2007}. For our purposes, this provides a
physically consistent and observationally plausible implementation of
galaxy formation within the dynamical framework of $\Lambda$CDM. It
can therefore be used to explore the differences between galaxy growth
and dark halo growth in this structure formation model. We use the
public database containing the properties of the dark halos and the
galaxies\footnote{http://www.mpa-garching.mpg.de/millennium} to
construct mean growth rates for galaxies through major mergers,
through minor mergers and through star-formation, each as a function
of galaxy mass and of redshift, and we compare these with analogously
defined growth rates for dark halos.

Our paper is organized as follows. In Sec.~\ref{sec:simu} we introduce the
\emph{Millennium Run} and the prescriptions used to simulate galaxy formation
using merger trees built from it. Sec.~\ref{sec:gstat} presents our analysis
of the mass and redshift dependence both of the major merger rate and of
growth rates through major and minor mergers as well as through star
formation. In Sec.~\ref{sec:Fstat}, we discuss the corresponding properties of
dark halos (defined here as FOF groups) and contrast them with our results for
galaxies.  Conclusions and discussions are presented in Sec.~\ref{sec:summary}.

% %%%%%%%%%%%%%%%%%%%%%%%%%%%%%%%%%%%%%%%%%%%%%%%%%%%%%%%%%%%%%%%%%%%%%%%%%%%%%

\begin{figure}
\bc
\hspace{-0.6cm}
\resizebox{8.5cm}{!}{\includegraphics{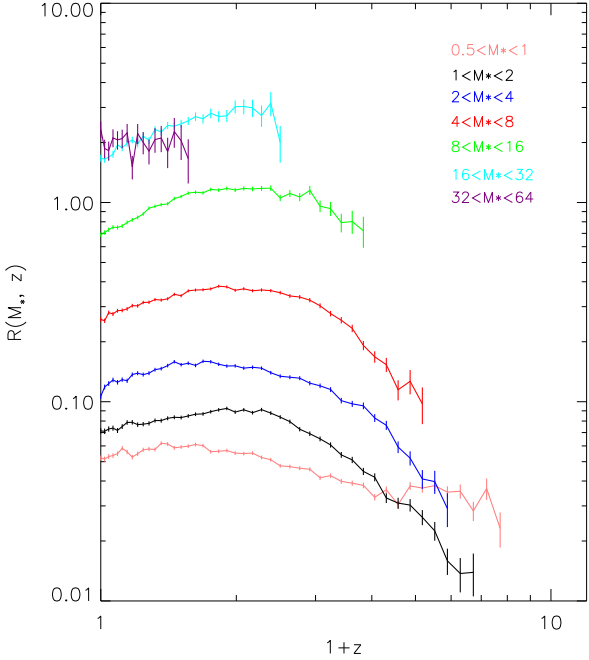}}\\%
\caption{ The specific rate of formation of galaxies through major mergers as
a function of redshift. The 7 curves refer to product galaxies with stellar
mass in 7 disjoint ranges, identified by labels with units of
$10^{10}M_\odot$. A galaxy is considered to have just formed through a major
merger if its two largest progenitors in the preceding Millennium Run output
differ by less than a factor of 3 in stellar mass. The dimensionless measure
of merger rate used here is the fraction of all galaxies in the given stellar
mass bin at redshift $z$ which form through a major merger per unit time,
multiplied by the age of the Universe at redshift $z$. Error bars give Poisson
uncertainties derived from the number of new merger products in each bin at
each redshift. The probability that a galaxy has just formed through a major
merger is a strong function of stellar mass, but a weak function of redshift.}

\label{fig:R}
\ec
\end{figure}

\section{The simulation and the galaxy formation model}
\label{sec:simu}
The galaxy catalogue used in this paper was produced using a ``hybrid
" technique: a large N-body simulation was first carried out to define
the evolution of the dark matter distribution, and then a suite of
semi-analytic prescriptions was implemented in order to simulate the
formation and evolution of galaxies within a stored ``forest'' of
(sub)halo merging trees constructed from the original simulation.  A
detailed description of the \emph{Millennium Simulation} and of the
galaxy formation model can be found in
\citep{Nature2005,Croton2006,LB2007}. Here we summarize the main
simulation characteristics and the way the halo merger trees were
constructed, as well as those aspects of the galaxy formation
modelling that are relevant to our study of galaxy growth.

\subsection{The simulation}
The \emph{Millennium Simulation} \citep{Nature2005} used in this study is the
largest simulation of cosmic structure formation so far carried out.  It
adopts the concordance $\Lambda$CDM cosmology and follows $N=2160^3$
particles from redshift $z=127$ to $z=0$ in a comoving box of side-length
$500h^{-1}$ Mpc. This volume is large enough to investigate rare objects such
as quasars and rich clusters of galaxies, yet, with $N=2160^3$ particles, has
a dark matter particle mass of only $8.6*10^{8}M_\odot$, allowing the galaxy
formation model to follow the formation of all galaxies more massive than the
Small Magellanic Cloud. The assumed cosmological parameters are $\Omega_{\rm
m}=0.25$, $\Omega_{\rm b}=0.045$, $h=0.73$, $\Omega_\Lambda=0.75$, $n=1$, and
$\sigma_8=0.9$, where the Hubble constant is parameterized as usual as $H_0=
100hkms^{-1}Mpc^{-1}$. These parameters are consistent with a combined
analysis of the 2dFGRS \citep{Colless2001} and the first-year WMAP data
\citep{Spergel2003}.

During the simulation, the full particle data were stored at 64 output
times approximately logarithmically spaced from $z=20$ until $z=1$ and
at approximately 200 Myr intervals thereafter. At each time, the
simulation code produced a friends-of-friends group catalogue on the
fly by linking together particles separated by less than 0.2 of the
average interparticle separation \citep{Davis1985}. Subsequently, the
SUBFIND algorithm \citep{Springel2001b} was used to divide each FOF
group into a disjoint set of self-bound subhalos. These subhalos are
the basis for the merger trees, which are defined by linking each
subhalo from a given output time to one and only one descendent at the
following output time.  When studying the growth of dark halos in
Sec.~\ref{sec:Fstat}, we define a halo as an FOF group and we estimate
its mass as the sum of the masses of all its subhalos. This typically
loses a small amount of ``diffuse'' material which was bound to none
of the subhalos. This is not significant for our purpose here, and
this definition was convenient, since the original FOF halo mass was
not stored in the (sub)halo database when we carried out this
project. More importantly, this mass definition allows us to deal in a
straightforward way with the problem that simulated halos, unlike our
simulated galaxies or the halos considered in simplified models for
halo growth, not only merge but can also fragment. Mass from a single
FOF halo can thus contribute to several FOF halos at some later time.

\label{sec:simulation}
\subsection{Merger rates}
 
In the galaxy formation models implemented on the Millennium Simulation, a
galaxy begins to condense at the centre of a halo as soon as it is identified
as a persistent object with more than 20 dark matter particles. As the halo
grows, so does the galaxy at its centre, forming stars at a rate governed by
its cold gas content and by empirically determined star formation ``laws''.
The halo may merge into a larger system, becoming an independent subhalo
orbiting within the FOF group. The galaxy is now considered a satellite, 
losing its supply of fresh gas, and perhaps ceasing to form stars if
it uses up its available interstellar medium. Dynamical friction effects bring
the orbit of the subhalo ever closer to the centre of its parent, and
tidal effects strip away its outer regions until eventually it may be disrupted
entirely (or at least drop below the resolution limit of the simulation). At
this point the galaxy is associated to the most-bound particle of the
subhalo at the last time it was identified and is marked as a candidate for
merging with the central galaxy of the parent halo. The merger takes
place one estimated dynamical friction time later.

Galaxy mergers may trigger strong star formation. In the galaxy
formation model of \cite{Croton2006} and \cite{LB2007} which we
analyse here, a recipe similar to that of \cite{Somerville2001} is
adopted to describe starbursts during minor mergers. In this model, a
fraction $e_{burst}$ of the cold gas of final galaxy is converted into
stars, where \begin{displaymath}
e_{burst}=0.56*(\frac{M_{satellite}}{M_{central}})^{0.7}.
\end{displaymath} A major merger is assumed to occur whenever the two
galaxies differ by a factor of less than 3 in baryonic mass. In such a
merger the starburst is assumed to convert a large fraction of the
cold gas into stars and to eject the rest from the galaxy. The remnant
of such a merger is assumed to be an elliptical galaxy. It may,
however, grow a new disk if gas is able to cool from the surrounding
halo, and in this case the merged system becomes the bulge of a larger
spiral galaxy.

In this galaxy formation model, central galaxies are treated differently than
satellites.  Only central galaxies are fed new material by cooling from the
hot atmosphere of their halo, by direct infall of cold gas, or by merging of
satellites.  No new material accretes onto satellite galaxies, so that their
star formation terminates when their cold gas is used up. Gas accretion
processes depend strongly on time and on galaxy mass. At early times and in
low-mass galaxies gas cools substantially more efficiently than in high-mass
systems and at late times. In addition, an important innovation in the model
of \cite{Croton2006} (and included here) is a treatment of ``radio mode''
feedback. This assumes that if the central galaxy has a supermassive black
hole and sits at the centre of a static hot atmosphere, then radio activity
will prevent further cooling of hot gas. This resolves the long-standing
``cooling flow problem'' and ensures that a massive elliptical at the centre
of a group or cluster does not grow a new disk and so remains ``red and
dead''. As a result the only significant growth mode for high mass galaxies
is through merging.
 
In this study, we consider all galaxies in the Millennium/DeLucia database
with stellar mass between $5\times 10^9M_\odot$ and $6.4\times
10^{11}M_\odot$. Although the galaxy catalogues are nearly complete to a mass
at least 5 times lower than this, we want to be able to resolve the recent
merging history of each system and so we adopt this more conservative
limit. This choice leaves us with a total 81896686 galaxies (summed over all
redshifts). To investigate the mass dependence of galaxy growth, we divide
this sample into seven mass bins, each a factor of 2 wide. The highest mass
bin contains the smallest number of galaxies, a total of 22827 systems.

%%%%%%%%%%%%%%%%%%%%%%%%%%%%%%%%%%%%%%%%%%%%%%%%%%%%%%%%%%%%%%%%%%%%%%%%%%%%%%%%%%%%%%%%%%   
\begin{figure}
\bc
\hspace{-0.6cm}
\resizebox{8.5cm}{!}{\includegraphics{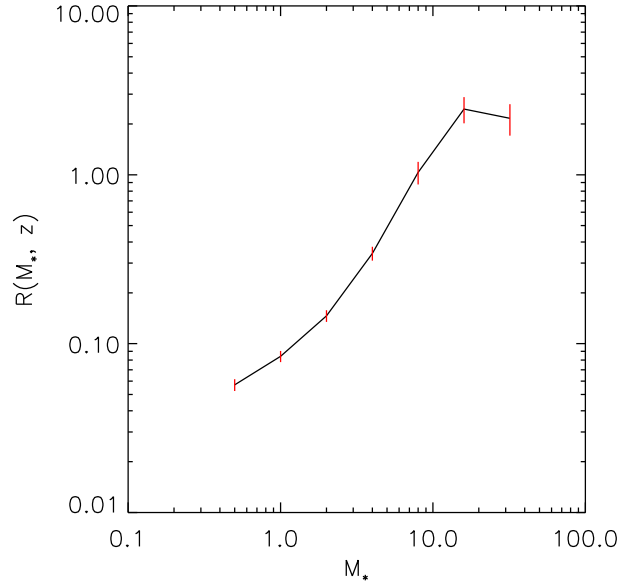}}\\%
\caption{ The relation between the stellar mass of galaxies and their
specific formation rate through major mergers. The rates given here
average the data plotted in Figure 1 over the redshift range $0\leq z
\leq 2$ (except for the highest stellar mass bins where there are
insufficient objects to determine a rate at the higher redshifts). The
error bars indicate the {\it rms} fluctuation in rate over these
redshift intervals. Clearly, the probability that a galaxy has just
formed through a major merger increases approximately linearly with stellar mass in this galaxy formation simulation.}
\label{fig:Rm}
\ec
\end{figure}

\section{Galaxy growth rates}
Growth in the stellar mass of galaxies occurs through two processes:
conversion of gas into stars (either quiescently or in a starburst)
and the addition of stars through mergers. In this section we mine the
publicly available database to study the interplay between these processes. We
begin by studying how the rate of major mergers depends on the mass of
the product galaxy and on redshift. We then compare mean galaxy growth
rates due to this process to mean growth rates due to all mergers
(major and minor) and to star formation. For each galaxy in the
database at each time, we define the main progenitor at the previous
stored time to be the progenitor with the largest stellar mass. If a
galaxy has more than one progenitor at the earlier time, then it has
undergone a merger between the two times. If $m$ of the other
progenitors differ from the main progenitor by less than a factor of 3
in stellar mass, then the galaxy is assumed to have had $m$
major mergers in this time interval.  \label{sec:gstat}
 
We define a dimensionless major merger rate per galaxy as a function of 
redshift and stellar mass through 
     \begin{equation}
R(M_*,z)=\frac{N_{major}(M_*,z)/\delta t(z)}{N_{gal}(M_*,z)/t(z)} 
 	\end{equation}
\label{Requ}
where $N_{gal}(M_*,z)$ is the number of galaxies in the simulation at
redshift $z$ and with stellar mass in a chosen interval centred on
$M_*$, $N_{merger}(M_*,z)$ is the number of these galaxies which have
had a major merger since the last stored redshift $z_p(z)$ (a galaxy
which has had $m$ major mergers is counted $m$ times), $\delta t(z)$
is the time interval between $z_p$ and $z$ and $t(z)$ is the age of
the universe at $z$. Hence $R(M_*, z)$ is the fraction of galaxies of
stellar mass $M_*$ formed per Hubble time through major mergers at
redshift $z$.

Fig.~\ref{fig:R} shows major merger rates estimated in this way as a
function of redshift for seven intervals of stellar mass, each a
factor of 2 wide.  We plot Poisson errors on our estimates which are
determined entirely by the number of merger remnants
$N_{major}(M_*,z)$ found at each time.  At low redshift ($z<2$) our
dimensionless rate depends remarkably weakly on redshift. For large
stellar mass, any variation is within the noise. For smaller $M_*$
there is a gentle rise (by less than a factor of 2) from low redshift
out to $z\sim 1.5$, followed by a decline at higher redshift. On the
other hand the dependence of $R(M_*,z)$ on stellar mass is very
strong.  The probability of formation through major mergers is about
40 times higher for the most massive galaxies we consider than for the
least massive galaxies.  Galaxies comparable in mass to our Milky Way form
through major mergers at a rate of about $25\%$ of the population per
Hubble time, while for galaxies with a stellar mass $\sim 4\times
10^{11}M_\odot$ the corresponding rate is about 8 times higher.
  
To see more clearly the stellar mass dependence of the specific rate
of formation through major mergers, we plot in Fig.~\ref{fig:Rm} the
relation between $\langle R(M_*,z)\rangle$ and stellar mass. Here we
have averaged the data of Fig.~\ref{fig:R} over the redshift interval
from $z=2$ to 0 (or, for massive galaxies, over redshifts where there
are more than 15 major mergers in total). Error bars show the {\it
rms} variation in the rate over the redshift range used.  The relative
formation rate through major mergers is approximately proportional to
stellar mass $R\propto M_*$, although the plot suggests a more complex
behaviour with an initial steepening towards higher mass followed by a
(possible) saturation at the highest mass.
\begin{figure*}
\bc
\hspace{-1.6cm}
\resizebox{17.cm}{!}{\includegraphics{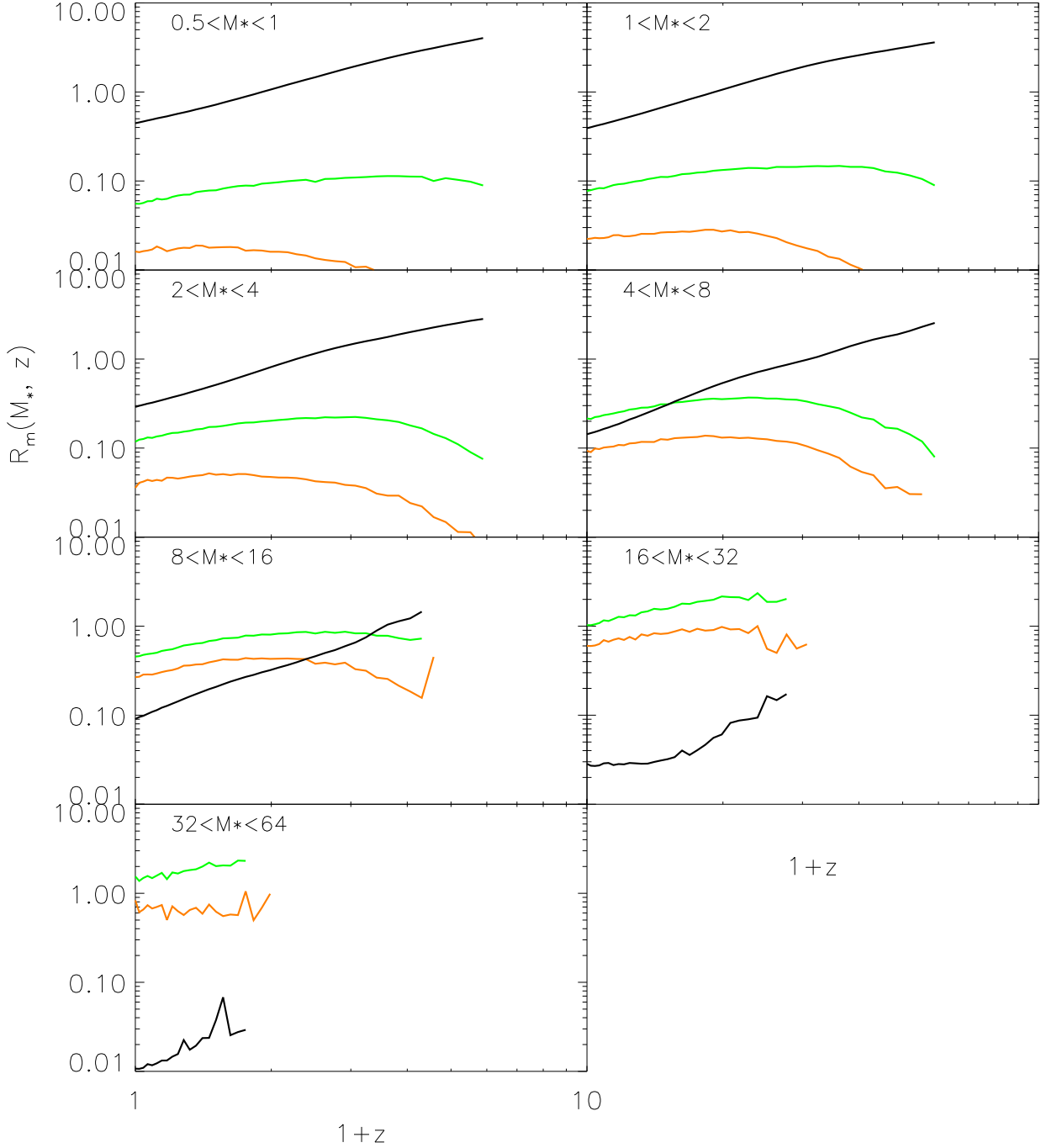}}\\%
\caption{ Dimensionless mean growth rates for galaxies as a function
of redshift for the 7 different stellar mass bins of Figure 1 and for
different growth modes. For each galaxy at each output time, the
fraction of its stellar mass gained in a particular mode since the
previous output is divided by the time between outputs and multiplied
by the current age of the Universe.  The result is then averaged over
all galaxies in the chosen mass bin and plotted against output
redshift. The different curves represent stellar mass growth through
major mergers (orange) through all mergers (green) and through star
formation (black). The stellar mass ranges in the labels for each panel are
given in units of $10^{10}M_\odot$.}
\label{fig:GR}
\ec
\end{figure*}

Galaxies grow not only through major mergers, but also through minor mergers
and through star formation. In order to compare the relative importance of
these processes, we now calculate mean growth rates for galaxies in each of
these channels as a function of stellar mass and redshift. In analogy to
equation (1) we define mean dimensionless growth rates due to major
mergers, to all mergers and to star formation as
     \begin{equation}
R_{m,major}(M_*,z)=\frac{M_{major}(M_*,z)/\delta t(z)}{M_{gal}(M_*,z)/t(z)} 
 	\end{equation}
\label{Rmajor}
  \begin{equation}
R_{m,merger}(M_*,z)=\frac{M_{merger}(M_*,z)/\delta t(z)}{M_{gal}(M_*,z)/t(z)} 
 	\end{equation}
\label{Rmerger}
     \begin{equation}
R_{m,gas}(M_*,z)=\frac{M_{gas}(M_*,z)/\delta t(z)}{M_{gal}(M_*,z)/t(z)} 
 	\end{equation}
\label{Rgas}
where $M_{gal}(M_*,z)$ is the total stellar mass of all galaxies at
redshift $z$ with individual stellar masses in the bin centred on
$M_*$, and $M$ with subscripts `major', `merger' and `gas' indicates
the total stellar mass added to the main progenitors of these galaxies
since the previous output time through major mergers, all mergers and
star formation, respectively. This includes star formation over this
time interval in all the progenitor galaxies, as well as in quiescent
and in merger-related starburst modes.  $\delta t(z)$ and $t(z)$
have the same meaning as before.  These rates represent the recent
growth of galaxies {\it prior} to the time they are observed in terms
of the fractional increase in their stellar mass per current Hubble
time occurring in each of the three modes. For example, $R_{m,gas} >1$
represents a class of galaxies whose recent average star formation
rate exceeds their past average star formation rate.
\begin{figure}
\bc
\hspace{-0.6cm}
\resizebox{8.5cm}{!}{\includegraphics{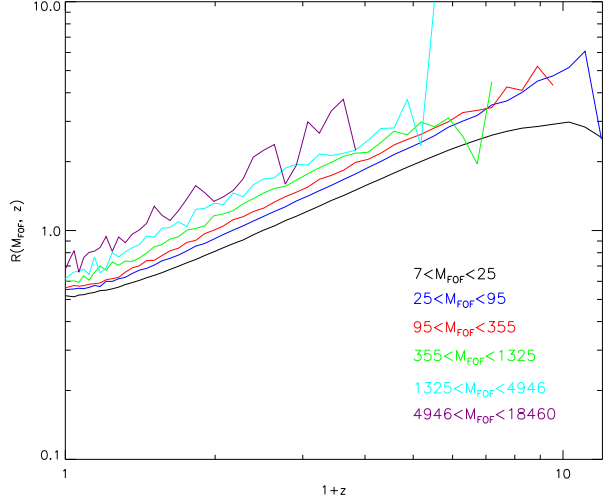}}\\%
\caption{ For comparison with Fig.~\ref{fig:R} we plot redshift against the
specific formation rate of FOF groups through major mergers, averaged
over groups in 6 different mass bins as indicated by label color.  The
mass unit here is $10^{10}M_\odot$.}
\label{fig:RH}
\ec
\end{figure}
     
In Fig. ~\ref{fig:GR} we plot these growth rates as a function of
redshift for the same 7 bins of stellar mass already illustrated in
Fig.~\ref{fig:R}.  The orange curves give the dimensionless growth
rate through major mergers and so are very similar to the curves
already plotted in Fig.~\ref{fig:R}.  Indeed, the ratio of the two is
just the average of the ratio of the stellar mass of the smaller
galaxy in a major merger to the stellar mass of the merger
product. Thus, the dimensionless growth of galaxies through major
mergers also depends little on redshift but strongly on stellar mass
(as in Fig.~\ref{fig:Rm}). Only for the most massive galaxies does
$R_{major}$ approach unity; for galaxies of Milky Way mass it is
around 10\% at all redshifts.

The green curves in Fig.~\ref{fig:GR} give mean growth rates due to
all mergers. For all but the more massive galaxies at the lowest
redshifts, these curves lie more than a factor of 2 above the major
merger curves.  The difference between the two curves increases with
increasing redshift in all cases.  Thus, minor mergers are generally
{\it more} important for increasing the stellar mass of galaxies than
are major mergers. For small mass galaxies at high redshift the ratio
of the two growth rates can be an order of magnitude. For galaxies
with masses above $10^{11}{\rm M}_\odot$ (in the model this
represents the classical giant elliptical population) merging
dominates the growth rates at redshifts $z<2$, and major mergers
account for more than half of the total stellar mass growth at low
redshifts. In the highest stellar mass bin the relative importance of
major and minor mergers is slightly different; these objects are the
Brightest Cluster Galaxies investigated in detail by De Lucia \&
Blaizot (2007).

Finally, the black curves in Fig.~\ref{fig:GR} give mean growth rates
due to star formation as a function of redshift. These are constructed
by averaging {\it all} the star formation between two output times in
{\it all} the progenitors of the galaxies in each mass bin. As a
result, they include quiescent star formation both in the main
galaxies and in smaller galaxies which merge with them, as well
merger-induced starbursts. Unlike the growth rates due to mergers,
they increase monotonically and relatively steeply towards high
redshift, roughly as one power of $(1+z)$ on average, although the
slope decreases with redshift at low stellar mass and increases with
redshift at high stellar mass. At the present day $R_{m,gas}(M_*,0)$
is a decreasing function of $M_*$ and is always below unity.  Thus,
galaxies of all stellar masses are, on average, currently forming
stars at less than their past average rate. For galaxies of Milky Way
mass, the mean star formation rate at $z=0$ is about 15\% of the past
average; this ratio drops to very small values for more massive
systems.  

This behaviour is well known in the real Universe and is often taken
as evidence for ``downsizing''; massive galaxies seem to complete most of
their star formation at higher redshift than low mass systems.
Somewhat surprisingly, however. this ranking of dimensionless growth
rate holds at {\it all} redshifts, not just at $z=0$. In this model
there is {\it no} redshift at which high stellar mass galaxies are
growing faster (in relative terms) than less massive systems. Except
for the highest mass bin (where galaxies form almost exclusively
through multiple mergers) the dimensionless growth rates due to star
formation exceed unity at sufficiently high redshift for galaxies of
all stellar mass. This remains true to lower redshift for lower
stellar mass.

If we compare the mean growth rates due to star formation with those
due to (all) mergers, we see that, except at the highest stellar
masses, star formation dominates at sufficiently high redshift.  This
is true all the way down to $z=0$ for galaxies less massive than the
Milky Way, but for higher mass systems mergers are the dominant growth
mode at low redshift. It is interesting that the Milky Way mass, which
is also approximately the characteristic stellar mass at the knee of
the galaxy luminosity function, marks the boundary between the two
regimes. This is not a coincidence. It is built into the model by the
physical assumptions required to get a good fit to the observed galaxy
luminosity function. In low-mass systems cooling is very efficient and
supernova feedback has to be invoked to prevent overproduction of
stars.  Even with such feedback, a significant fraction of the
baryonic material gained by small halos is turned into stars, and most
of this accreted material is associated with objects which were too
small to contain stars of their own.  Hence star formation is a more
effective growth mode than merging. At Milky Way mass, cooling is
still efficient, particularly at early times, and supernova feedback
is less effective in preventing star formation.  On the other hand,
much of the infalling material is in objects which are massive enough
to contain substantial numbers of their own stars.  Thus stellar
mergers become competitive with star formation.  For higher stellar
masses, the model invokes ``radio mode'' AGN feedback to suppress
cooling and star formation.  The steep quasi-exponential tail of the
stellar mass function is then populated almost exclusively by mergers.

%%%%%%%%%%%%%%%%%%%%%%%%%%%%%%%%%%%%%%%%%%%%%%%%%%%%%%%%%%%%%%%%%%%%%%%%%%%%%%5
\begin{figure*}
\bc
\hspace{-1.6cm}
\resizebox{17.cm}{!}{\includegraphics{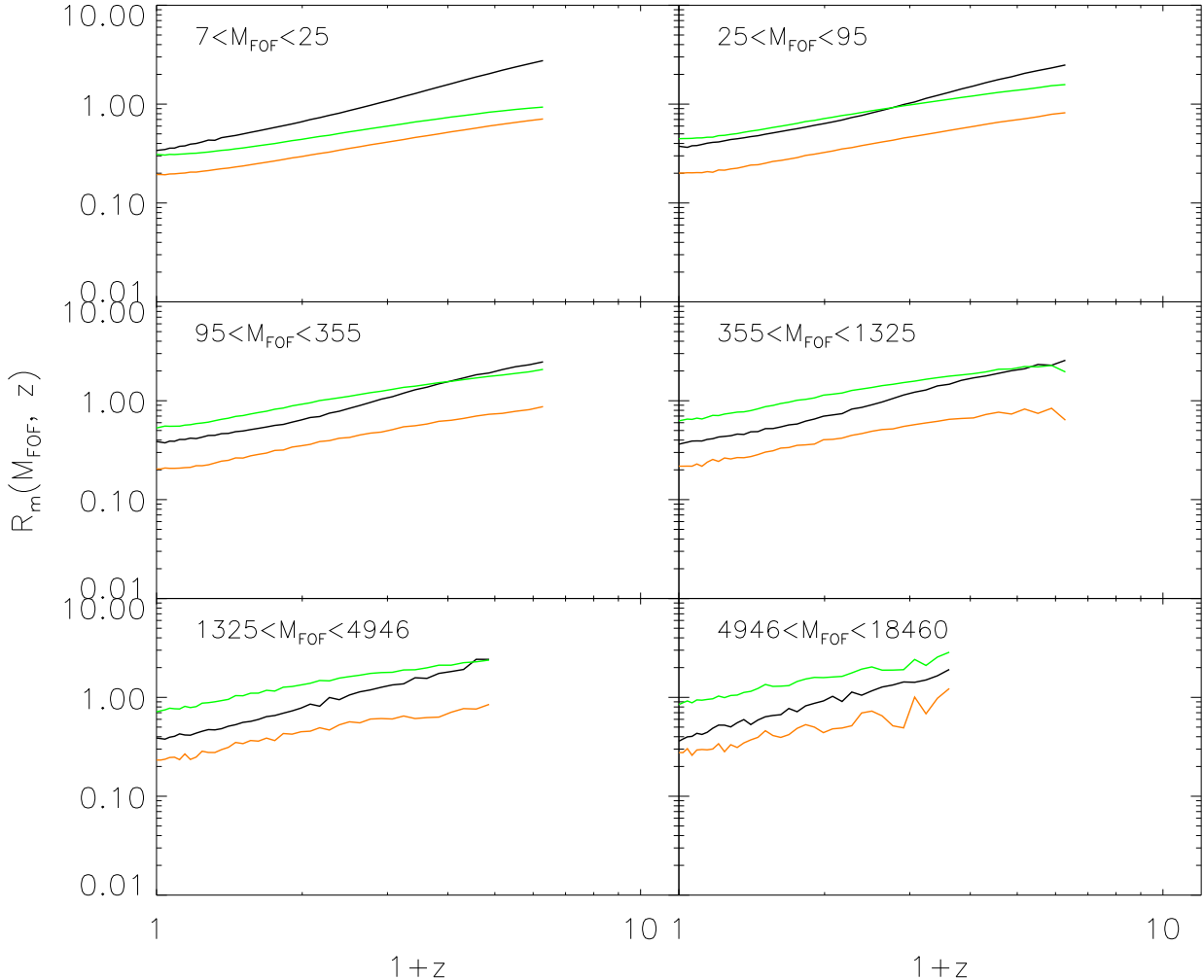}}\\%
\caption{For comparison with Fig.~\ref{fig:GR}, we plot dimensionless
mass accretion rates for FOF groups as a function of redshift for the
same 6 mass bins as in Fig.~\ref{fig:RH}. The different curves give
the mean mass accretion rate due to major mergers (orange), to all
mergers (green) and to accretion of diffuse particles (black).  The
mass unit for the labels in each panel is $10^{10}M_\odot$.}
\label{fig:HR}
\ec
\end{figure*}

\section{Growth Rates for FOF groups}
\label{sec:Fstat}
As discussed in Sec.~\ref{sec:intro}, the relation between galaxy
mergers and mergers of their host halos is less straightforward than
one might expect. In this section we investigate merger and growth
rates for dark halos in a way which allows direct comparison with the
results presented for galaxies above.  For the purposes of our study
it is convenient to identify dark halos as the friends-of-friends
(FOF) groups initially identified in the Millennium Simulation, and to
approximate the mass of each FOF group by the sum of the masses of its
identified subhalos. This loses the mass of a certain number of
``diffuse'' particles which are not bound to any subhalo, but this
systematic is relatively small for most halos and is of no consequence
for our analysis.  This scheme provides a straightforward way for us
to deal with the problem that simulated halos, unlike those in
extended Press-Schechter models (e.g. Lacey \& Cole 1993) or the
galaxies discussed above, often fragment into pieces which can become
parts of different halos at a later time. This means that the
progenitors of an FOF halo may include only part of an earlier FOF
halo.  Tracking individual subhalos allows us to account for this,
since the Millennium halo database is set up so that each subhalo has
a unique descendent, ensuring that the progenitors of an FOF group are
a unique set of subhalos which may form all or part of several FOF
halos.

We bin the FOF groups at each redshift according to mass, with each
bin spanning a factor of 3.8 in mass. The six bins for which we
present results then correspond very roughly to halos whose central
galaxies lie in the upper six stellar mass bins of figures \ref{fig:R}
and \ref{fig:GR}.

In Fig~\ref{fig:RH}, we plot the redshift dependence of the specific rate of formation of FOF
halos through major mergers for our 6 bins of halo
mass. A halo is defined to have just undergone $m$ major mergers if
its progenitor subhalos at the previous output come from at least
$m+1$ different FOF halos, and if the total subhalo mass coming from
$m$ of the subdominant FOF progenitors is more than a third of that
coming from the main FOF progenitor. This merger count can then be
used to define a merger rate in direct analogy to equation
(1). Figure~\ref{fig:RH} can be compared directly with the 6 most
massive cases of Fig.~\ref{fig:R}. The behaviour is quite different,
however. In Fig.~\ref{fig:RH} there is a strong and monotonic
dependence of formation rate on redshift, but there is little
dependence on halo mass.  This is the exact contrary of what we found
for galaxies, where the mass dependence was strong and the redshift
dependence weak.  The redshift dependence of these curves is
reasonably well described as a simple proportionality to $(1+z)$. For
all masses the rates exceed unity for all but the lowest
redshifts. Recall that in Fig.~\ref{fig:R} we found the coresponding
rates for galaxies to exceed unity only for the most massive
systems. Major mergers are thus a much more significant growth mode
for dark halos than they are for most galaxies.

Fig ~\ref{fig:HR} shows dimensionless growth rates for FOF halos as a
function of redshift for the same 6 halo mass bins. These rates are defined in
exact analogy to equations (4) through (2) and refer to
growth through major mergers (orange), through all resolved mergers (green),
and through accretion of ``diffuse'' particles (i.e. simulation particles not
assigned to any FOF halo with more than 20 particles; the black curve).  Again
the growth rate through major mergers parallels the specific formation rate
already plotted in Fig.~\ref{fig:RH}; the ratio of the two is just the average
mass of the smaller partner in a major merger in units of the final halo mass.
Both the growth rate through major mergers and the growth rate through all
(resolved) mergers are near power-laws of similar slope.  The growth rate
through all resolved mergers exceeds that through major mergers by a larger
factor for high-mass halos than for low-mass ones. This primarily reflects the
fact that the resolution limit of the simulation corresponds to a much lower
mass ratio limit for identifying a merger in the former case. This is not the
whole story, however, as one can see by the fact that the diffuse accretion
rate depends differently on redshift than the other growth rates.  Hence the
growth of objects of {\it given} mass is more strongly affected by accretion
of diffuse material at early times than at late times.  In addition, comparing
the major merger growth rates (the orange curves) with the total growth rates
(the sum of the green and black curves), one sees that while at high masses
and at early redshifts major mergers account for about 15\% of the total
growth rate, for small objects at late times they account for a larger
fraction of the growth. Note that at all redshifts and for all masses,
accretion of ``diffuse'' particles accounts for at least 30\% of the total
growth.

%%%%%%%%%%%%%%%%%%%%%%%%%%%%%%%%%%%%%%%%%%%%%%%%%%%%%%%%%%%%%%%%%%%%%%%%%%%%%%%%%%%%%%%%%%   
\section{Summary and discussion}
\label{sec:summary}

We have used publicly available data for the \cite{LB2007} model of
galaxy formation to study the relative importance of merging and of
star formation for the growth of galaxies. This model is based on
stored halo merging trees for the \emph{Millennium Simulation}, a very
large simulation of the evolution of the dark matter distribution in a
$\Lambda$CDM cosmology. It is consistent with a wide variety of
observational data on the properties and clustering of galaxies both
at low and at high redshift. We thus expect its behaviour to give at
least a qualitative indication of the balance needed between the
various modes of galaxy assembly in any successful model in the
$\Lambda$CDM context. A particular goal of our study has been to
contrast the roles of merging in galaxy and dark halo evolution.

The most striking result from our study is that formation through
merging depends in a completely different way on mass and redshift
for our two classes of object. Recent formation through a major merger
is almost equally likely for halos of all masses at any given time,
but is substantially more likely at early times than it is today. For
galaxies, on the other hand, the likelihood of recent formation
through a major merger is a strong function of stellar mass, but
depends at most weakly on redshift. In addition, halos of all masses
have grown more rapidly through mergers than all but the most massive
galaxies. A little reflection shows that these differences are
required by the facts that a galaxy cluster is considered as a single
dark matter halo but contains many distinct galaxies, and that the
stellar mass function for cluster galaxies differs little from that of the
Universe as a whole.  This implies that the build-up of massive halos
through mergers cannot be paralleled by merging of the associated
galaxies. Merging plays a much less important role (though still
significant) in galaxy growth than in dark halo growth. The high rates
of recent merging found for the most massive galaxies are a selection
effect. Only through merging can galaxies attain such high
masses. This is also the reason why the most massive galaxies are
usually ellipticals,

A second striking result from our study is the increasing importance
of star formation with increasing redshift for galaxies of all
masses. At low redshift we find the observed result that mean specific
growth rates through star formation are smaller in high-mass galaxies
than in low-mass ones, but it turns out that this result also holds at
high redshift.  According to the \cite{LB2007} model there is no
redshift where the specific star formation rate of massive galaxies
significantly exceeds that of low mass systems. Individual objects may
be experiencing dramatic starbursts, but averaged over the population
of all objects of given stellar mass, the prediction is that the mean
specific growth rate through star formation is always a decreasing
function of stellar mass. 

Only at redshifts below one and for galaxies
comparable to or more massive than the Milky Way does the growth rate
through mergers exceed that through star formation. This corresponds
nicely to the ``transition stellar mass'' at which the stellar
populations and the structural parameters of local galaxies switch
from being predominantly star-forming and disk-like to predominantly
old and spheroidal \citep{Kauffmann2003}.  This agreement is, of
course, in part a consequence of the tuning of the parameters of the
galaxy formation model to fit observation.

A less surprising but still interesting result is that merger-related
growth for objects of all stellar masses and at most times is roughly equally divided between what we have defined as major and
minor mergers.  Clearly our separation at a progenitor stellar mass
ratio of 3 to 1 is arbitrary.  If we had chosen 5 to 1, major mergers
would have dominated in most cases.  If we had chosen 50 to 1, minor
mergers would have been unimportant. Clearly the accretion of the LMC
will make a much more significant change to the Milky Way's stellar
mass than the addition of all the Dwarf Spheroidals, and this in turn
will be dwarfed by the impending merger with M31!

As we now show, the FOF halo behaviour we find is at least
qualitatively consistent with the predictions of EPS theory
\citep{LC1993}. The analytical expression for the probability that a mass
element which is part of a halo of given mass $M_2$ at time $t_2$ is
part of a halos of (smaller) mass $M_1$ at the earlier time
$t_1$ is
\begin{displaymath}
f(S_1,\omega_1|S_2,\omega_2)dS_1
\end{displaymath}
\begin{displaymath}
=\frac{\omega_1-\omega_2}{(2\pi)^{1/2}(S_1-S_2)^{3/2}}\exp[-\frac{(\omega_1-\omega_2)^2}{2(S_1-S_2)}]dS_1
\end{displaymath} 
where $S_{1,2}$ is the {\it rms} linear density fluctuation
(extrapolated to $z=0$) in spheres containing a mean mass $M_{1,2}$,
$\omega_{1,2}\equiv \delta_{c0}/D(z_{1,2})$ is the redshift-dependent
critical density for collapse, $D(z)$ is the growth factor of linear
fluctuations, and $\delta_{c0}\approx 1.69$ is a constant. By taking
the limit as $t_2$ tends to $t_1$ (so $\omega_2-\omega_1$ tends to 0)
and integrating over $S_1$, we can get the dimensionless merger rate
per product halo:
\begin{displaymath}
\frac{P(M_{high},M_{low},\omega_1|S_2,\omega_2)}{dt/t} 
\end{displaymath}
\begin{displaymath}
=\frac{1}{2}\int_{S(M_{low})}^{S(M_{high})}\frac{M_2}{M_1}\frac{t*d\omega/dt}{(2\pi)^{1/2}(S_1-S_2)^{3/2}}dS_1 
\end{displaymath}
Setting $M_{low}=\frac{M_2}{3}$ and $M_{high}=\frac{2M_2}{3}$ we get
the major merger rate which is seen to evolve with time as
$t*d\omega/dt$. In a Einstein de Sitter universe $D(z)\propto
(1+z)^{-1}\propto t^{2/3}$ and thus $t*d\omega /dt \propto (1+z)$,
roughly reproducing the behaviour we get for the major merger rate of
FOF halos in the Millennium Simulation. In the $\Lambda$CDM cosmology,
the formula is more complex but is quantitatively similar. As shown
by \cite{CPT1992}, $D(z)=g(z)/[g(0)(1+z)]$ where
\begin{displaymath}
g(z)\approx 5/2\Omega_m[\Omega_m^{4/7}-\Omega_{\Lambda}+(1+\Omega_m/2)(1+\Omega_{\Lambda}/70)]^{-1}
\end{displaymath}
and $\Omega_m$ ($\Omega_{\Lambda}$) is the density parameter of matter
(dark energy). We plot $t*d\omega/dt$ against $1+z$ for the
two cases in Fig 6 to illustrate the size of the expected differences.

The same formalism also allows the dimensionless mass accretion rates
through mergers and/or smooth accretion to be expressed as
\begin{displaymath}
\frac{P_M(M_{high},M_{low},\omega_1|S_2,\omega_2)}{dt/t}
\end{displaymath}
\begin{displaymath}
=\frac{1}{2}\int_{S(M_{low})}^{S(M_{high})}\frac{M_2}{M_1}\frac{t*d\omega/dt}{(2\pi)^{1/2}(S_1-S_2)^{3/2}}\frac{min(M_1,M_2-M_1)}{M_2}dS_1 
\end{displaymath}
The $M_2$ dependence of this rate can be seen by assuming the limits
$M_{low}$ and $M_{high}$ to scale with $M_2$, and approximating the
dependence of $S$ on $M$ as a power-law $S \propto M^{\alpha}$, where
$\alpha=-(n+3)/3$ for the usual definition of the density power
spectrum index $n$. The $rhs$ of the above equation then scales
as $M_2^{-\alpha/2}$. When $n$ lies in the expected range between $-2$
and $-1$, the mass dependence is very weak, roughly $\sim
M_2^{0.2}$. Taking into account that in the simulation one cannot
really take infinitesimal time intervals, this dependence on final
halo mass may be further weakened by the exponential term in the
expression for $f$.

Finally, we can also use these formulae to estimate the ratio
of the growth rate through major mergers to that through ``smooth''
accretion (here defined as $M_1 < \frac{1}{3}M_2$).
It is not easy to obtain analytic expressions for this ratio but it
can easily be computed from the above formulae in the power law
approximation for $S(M)$. Here we give  in Table 1
the relative fraction for several typical values of $\alpha$.
 \begin{table*}
 \caption{Table 1:Relative growth rates due to major mergers and to
other accretion modes.}

\begin{tabular}{||l||c||} 

\hline
$\alpha$ & $R_m(major mergers) : R_m(smooth accretion)$ \\
\hline
-2/3       &  0.27  : 0.73 \\
-1/3      &   0.26  : 0.74 \\
-1         &   0.29  : 0.71 \\
\hline
\end{tabular} 

\end{table*}
Roughly speaking, major mergers are predicted to contribute 25\% to
30\% of the total mass accretion, quite consistent with our simulation
results.

\begin{figure}
\bc
\hspace{-0.6cm}
\resizebox{8.5cm}{!}{\includegraphics{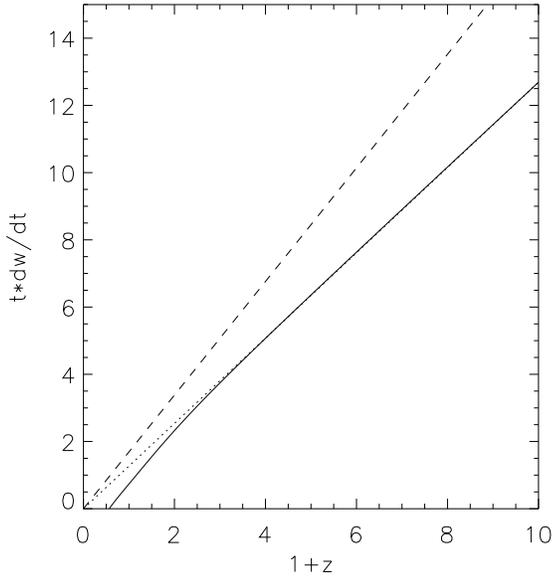}}\\%
\caption{Predicted relation between $t*d\omega/dt$ (proportional to
the dimensionless merger rate) and $1+z$ for the concordance
$\Lambda$CDM model (solid curve) and for an Einstein de Sitter
universe (dashed curve) according to extended Press-Schechter
theory. At redshifts above about 0.5 the two quantities are very nearly
proportional to each other in the $\Lambda$CDM case also, as shown by
the dotted straight line.}  \ec
\end{figure}
%%%%%%%%%%%%%%%%%%%%%%%%%%%%%%%%%%%%%%%%%%%%%%%%%%%%%%%%%%%%%%%%%%%

%%%%%%%%%%%%%%%%%%%%%%%%%%%%%%%%%%%%%%%%%%%%%%%%%%%%%%%%%%%%%%%%%%%
\section*{Acknowledgements}
The public databases on which this work is based can be found at
http://www.mpa-garching.mpg.de/millennium.  We are grateful to
Gabriella De Lucia, to Jeremy Blaizot and particularly to Gerard
Lemson for help in using the these databases, as well as to Shaun Cole
and Volker Springel for useful discussions.

\bibliographystyle{mn2e}

\bibliography{merger}

\begin{thebibliography}{}

\bibitem[\protect\citeauthoryear{{Abadi}, {Navarro}, {Steinmetz} \&
  {Eke}}{{Abadi} et~al.}{2003}]{Abadi2003}
{Abadi} M.~G.,  {Navarro} J.~F.,  {Steinmetz} M.,    {Eke} V.~R.,  2003, \apj,
  597, 21

\bibitem[\protect\citeauthoryear{{Barnes}}{{Barnes}}{1988}]{Barnes1988}
{Barnes} J.~E.,  1988, \apj, 331, 699

\bibitem[\protect\citeauthoryear{{Barnes} \& {Hernquist}}{{Barnes} \&
  {Hernquist}}{1991}]{BH1991}
{Barnes} J.~E.,  {Hernquist} L.~E.,  1991, \apjl, 370, L65

\bibitem[\protect\citeauthoryear{{Bell}, {Phleps}, {Somerville}, {Wolf},
  {Borch} \& {Meisenheimer}}{{Bell} et~al.}{2006}]{Bell2006}
{Bell} E.~F.,  {Phleps} S.,  {Somerville} R.~S.,  {Wolf} C.,  {Borch} A.,
  {Meisenheimer} K.,  2006, \apj, 625,270

\bibitem[\protect\citeauthoryear{{Benson}, {Cole}, {Frenk}, {Baugh} \&
  {Lacey}}{{Benson} et~al.}{2000}]{Benson2000}
{Benson} A.~J.,  {Cole} S.,  {Frenk} C.~S.,  {Baugh} C.~M.,    {Lacey} C.~G.,
  2000, \mnras, 311, 793

\bibitem[\protect\citeauthoryear{{Bower}, {Benson}, {Malbon}, {Helly}, {Frenk},
  {Baugh}, {Cole} \& {Lacey}}{{Bower} et~al.}{2006}]{Bower2006}
{Bower} R.~G.,  {Benson} A.~J.,  {Malbon} R.,  {Helly} J.~C.,  {Frenk} C.~S.,
  {Baugh} C.~M.,  {Cole} S.,    {Lacey} C.~G.,  2006, \mnras, 370, 645

\bibitem[\protect\citeauthoryear{{Bundy}, {Fukugita}, {Ellis}, {Kodama} \&
  {Conselice}}{{Bundy} et~al.}{2004}]{Bundy2004}
{Bundy} K.,  {Fukugita} M.,  {Ellis} R.~S.,  {Kodama} T.,    {Conselice} C.~J.,
   2004, \apjl, 601, L123

\bibitem[\protect\citeauthoryear{{Carlberg}, {Cohen}, {Patton}, {Blandford},
  {Hogg}, {Yee}, {Morris}, {Lin}, {Hall}, {Sawicki}, {Wirth}, {Cowie}, {Hu} \&
  {Songaila}}{{Carlberg} et~al.}{2000}]{Carlberg2000}
{Carlberg} R.~G.,  {Cohen} J.~G.,  {Patton} D.~R.,  {Blandford} R.,  {Hogg}
  D.~W.,  {Yee} H.~K.~C.,  {Morris} S.~L.,  {Lin} H.,  {Hall} P.~B.,  {Sawicki}
  M.,  {Wirth} G.~D.,  {Cowie} L.~L.,  {Hu} E.,    {Songaila} A.,  2000, \apjl,
  532, L1

\bibitem[\protect\citeauthoryear{{Carroll}, {Press} \& {Turner}}{{Carroll}
  et~al.}{1992}]{CPT1992}
{Carroll} S.~M.,  {Press} W.~H.,    {Turner} E.~L.,  1992, \araa, 30, 499

\bibitem[\protect\citeauthoryear{{Cen} \& {Ostriker}}{{Cen} \&
  {Ostriker}}{1992}]{CO1992}
{Cen} R.,  {Ostriker} J.~P.,  1992, \apj, 393, 22

\bibitem[\protect\citeauthoryear{{Cen} \& {Ostriker}}{{Cen} \&
  {Ostriker}}{2000}]{CO2000}
{Cen} R.,  {Ostriker} J.~P.,  2000, \apj, 538, 83

\bibitem[\protect\citeauthoryear{{Cimatti}, {Daddi} \& {Renzini}}{{Cimatti}
  et~al.}{2006}]{CDR2006}
{Cimatti} A.,  {Daddi} E.,    {Renzini} A.,  2006, \aap, 453, L29

\bibitem[\protect\citeauthoryear{{Cole}, {Aragon-Salamanca}, {Frenk}, {Navarro}
  \& {Zepf}}{{Cole} et~al.}{1994}]{Cole1994}
{Cole} S.,  {Aragon-Salamanca} A.,  {Frenk} C.~S.,  {Navarro} J.~F.,    {Zepf}
  S.~E.,  1994, \mnras, 271, 781

\bibitem[\protect\citeauthoryear{{Cole}, {Lacey}, {Baugh} \& {Frenk}}{{Cole}
  et~al.}{2000}]{Cole2000}
{Cole} S.,  {Lacey} C.~G.,  {Baugh} C.~M.,    {Frenk} C.~S.,  2000, \mnras,
  319, 168

\bibitem[\protect\citeauthoryear{{Colless}, {Dalton}, {Maddox}, {Sutherland},
  {Norberg}, {Cole}, {Bland-Hawthorn}, {Bridges}, {Cannon}, {Collins}, {Couch},
  {Cross}, {Deeley}, {De Propris}, {Driver}, {Efstathiou}, {Ellis}, {Frenk},
  {Glazebrook}, {Jackson}, {La}}{{Colless} et~al.}{2001}]{Colless2001}
{Colless} M.,  {Dalton} G.,  {Maddox} S.,  {Sutherland} W.,  {Norberg} P.,
  {Cole} S.,  {Bland-Hawthorn} J.,  {Bridges} T.,  {Cannon} R.,  {Collins} C.,
  {Couch} W.,  {Cross} N.,  {Deeley} K.,  {De Propris} R.,  {Driver} S.~P.,
  {Efstathiou} G.,  {Ellis} R.~S.,  {Frenk} C.~S.,  {Glazebrook} K.,  {Jackson}
  C.,  {Lahav} O.,  {Lewis} I.,  {Lumsden} S.,  {Madgwick} D.,  {Peacock}
  J.~A.,  {Peterson} B.~A.,  {Price} I.,  {Seaborne} M.,    {Taylor} K.,  2001,
  \mnras, 328, 1039

\bibitem[\protect\citeauthoryear{{Conselice}, {Bershady}, {Dickinson} \&
  {Papovich}}{{Conselice} et~al.}{2003}]{Conselice2003}
{Conselice} C.~J.,  {Bershady} M.~A.,  {Dickinson} M.,    {Papovich} C.,  2003,
  \aj, 126, 1183

\bibitem[\protect\citeauthoryear{{Croton}, {Springel}, {White}, {De Lucia},
  {Frenk}, {Gao}, {Jenkins}, {Kauffmann}, {Navarro} \& {Yoshida}}{{Croton}
  et~al.}{2006}]{Croton2006}
{Croton} D.~J.,  {Springel} V.,  {White} S.~D.~M.,  {De Lucia} G.,  {Frenk}
  C.~S.,  {Gao} L.,  {Jenkins} A.,  {Kauffmann} G.,  {Navarro} J.~F.,
  {Yoshida} N.,  2006, \mnras, 365, 11

\bibitem[\protect\citeauthoryear{{Davis}, {Efstathiou}, {Frenk} \&
  {White}}{{Davis} et~al.}{1985}]{Davis1985}
{Davis} M.,  {Efstathiou} G.,  {Frenk} C.~S.,    {White} S.~D.~M.,  1985, \apj,
  292, 371

\bibitem[\protect\citeauthoryear{{De Lucia} \& {Blaizot}}{{De Lucia} \&
  {Blaizot}}{2007}]{LB2007}
{De Lucia} G.,  {Blaizot} J.,  2007, \mnras, 375, 2

\bibitem[\protect\citeauthoryear{{De Lucia}, {Springel}, {White}, {Croton} \&
  {Kauffmann}}{{De Lucia} et~al.}{2006}]{Lucia2006}
{De Lucia} G.,  {Springel} V.,  {White} S.~D.~M.,  {Croton} D.,    {Kauffmann}
  G.,  2006, \mnras, 366, 499

\bibitem[\protect\citeauthoryear{{Di Matteo}, {Springel} \& {Hernquist}}{{Di
  Matteo} et~al.}{2005}]{Matteo2005}
{Di Matteo} T.,  {Springel} V.,    {Hernquist} L.,  2005, \nat, 433, 604

\bibitem[\protect\citeauthoryear{{Fall}}{{Fall}}{1979}]{Fall1979}
{Fall} S.~M.,  1979, \nat, 281, 200

\bibitem[\protect\citeauthoryear{{Farouki} \& {Shapiro}}{{Farouki} \&
  {Shapiro}}{1982}]{FS1982}
{Farouki} R.~T.,  {Shapiro} S.~L.,  1982, \apj, 259, 103

\bibitem[\protect\citeauthoryear{{Gao}, {White}, {Jenkins}, {Stoehr} \&
  {Springel}}{{Gao} et~al.}{2004}]{Gao2004}
{Gao} L.,  {White} S.~D.~M.,  {Jenkins} A.,  {Stoehr} F.,    {Springel} V.,
  2004, \mnras, 355, 819

\bibitem[\protect\citeauthoryear{{Ghigna}, {Moore}, {Governato}, {Lake},
  {Quinn} \& {Stadel}}{{Ghigna} et~al.}{1998}]{GG}
{Ghigna} S.,  {Moore} B.,  {Governato} F.,  {Lake} G.,  {Quinn} T.,    {Stadel}
  J.,  1998, \mnras, 300, 146

\bibitem[\protect\citeauthoryear{{Hatton}, {Devriendt}, {Ninin}, {Bouchet},
  {Guiderdoni} \& {Vibert}}{{Hatton} et~al.}{2003}]{Hatton2003}
{Hatton} S.,  {Devriendt} J.~E.~G.,  {Ninin} S.,  {Bouchet} F.~R.,
  {Guiderdoni} B.,    {Vibert} D.,  2003, \mnras, 343, 75

\bibitem[\protect\citeauthoryear{{Helly}, {Cole}, {Frenk}, {Baugh}, {Benson} \&
  {Lacey}}{{Helly} et~al.}{2003}]{Helly2003}
{Helly} J.~C.,  {Cole} S.,  {Frenk} C.~S.,  {Baugh} C.~M.,  {Benson} A.,
  {Lacey} C.,  2003, \mnras, 338, 903

\bibitem[\protect\citeauthoryear{{Kang}, {Jing}, {Mo} \& {B{\"o}rner}}{{Kang}
  et~al.}{2005}]{Kang2005}
{Kang} X.,  {Jing} Y.~P.,  {Mo} H.~J.,    {B{\"o}rner} G.,  2005, \apj, 631, 21

\bibitem[\protect\citeauthoryear{{Katz}, {Weinberg} \& {Hernquist}}{{Katz}
  et~al.}{1996}]{KWH1996}
{Katz} N.,  {Weinberg} D.~H.,    {Hernquist} L.,  1996, \apjs, 105, 19

\bibitem[\protect\citeauthoryear{{Kauffmann} \& {Charlot}}{{Kauffmann} \&
  {Charlot}}{1998}]{KC1998}
{Kauffmann} G.,  {Charlot} S.,  1998, \mnras, 297, L23+

\bibitem[\protect\citeauthoryear{{Kauffmann}, {Colberg}, {Diaferio} \&
  {White}}{{Kauffmann} et~al.}{1999}]{Kauffmann1999}
{Kauffmann} G.,  {Colberg} J.~M.,  {Diaferio} A.,    {White} S.~D.~M.,  1999,
  \mnras, 303, 188

\bibitem[\protect\citeauthoryear{{Kauffmann}, {Heckman}, {White}, {Charlot},
  {Tremonti}, {Peng}, {Seibert}, {Brinkmann}, {Nichol}, {SubbaRao} \&
  {York}}{{Kauffmann} et~al.}{2003}]{Kauffmann2003}
{Kauffmann} G.,  {Heckman} T.~M.,  {White} S.~D.~M.,  {Charlot} S.,  {Tremonti}
  C.,  {Peng} E.~W.,  {Seibert} M.,  {Brinkmann} J.,  {Nichol} R.~C.,
  {SubbaRao} M.,    {York} D.,  2003, \mnras, 341, 54

\bibitem[\protect\citeauthoryear{{Kauffmann}, {Nusser} \&
  {Steinmetz}}{{Kauffmann} et~al.}{1997}]{KNS1997}
{Kauffmann} G.,  {Nusser} A.,    {Steinmetz} M.,  1997, \mnras, 286, 795

\bibitem[\protect\citeauthoryear{{Kauffmann}, {White} \&
  {Guiderdoni}}{{Kauffmann} et~al.}{1993}]{KGW1993}
{Kauffmann} G.,  {White} S.~D.~M.,    {Guiderdoni} B.,  1993, \mnras, 264, 201

\bibitem[\protect\citeauthoryear{{Kitzbichler} \& {White}}{{Kitzbichler} \&
  {White}}{2007}]{KW2007}
{Kitzbichler} M.~G.,  {White} S.~D.~M.,  2007, \mnras, 376, 2

\bibitem[\protect\citeauthoryear{{Lacey} \& {Cole}}{{Lacey} \&
  {Cole}}{1993}]{LC1993}
{Lacey} C.,  {Cole} S.,  1993, \mnras, 262, 627

\bibitem[\protect\citeauthoryear{{Le F{\`e}vre}, {Abraham}, {}, {Ellis},
  {Brinchmann}, {Schade}, {Tresse}, {Colless} \& {Crampton}}{{Le F{\`e}vre}
  et~al.}{2000}]{Le2000}
{Le F{\`e}vre} O.,  {Abraham} R.,  {} S.~J.,  {Ellis} R.~S.,  {Brinchmann} J.,
  {Schade} D.,  {Tresse} L.,  {Colless} M.,    {Crampton} D.,  2000, \mnras,
  311, 565

\bibitem[\protect\citeauthoryear{{Lin}, {Koo}, {Willmer}, {Patton},
  {Conselice}, {Yan}, {Coil}, {Cooper}, {Davis}, {Faber}, {Gerke},
  {Guhathakurta} \& {Newman}}{{Lin} et~al.}{2004}]{Lin2004}
{Lin} L.,  {Koo} D.~C.,  {Willmer} C.~N.~A.,  {Patton} D.~R.,  {Conselice}
  C.~J.,  {Yan} R.,  {Coil} A.~L.,  {Cooper} M.~C.,  {Davis} M.,  {Faber}
  S.~M.,  {Gerke} B.~F.,  {Guhathakurta} P.,    {Newman} J.~A.,  2004, \apjl,
  617, L9

\bibitem[\protect\citeauthoryear{{Mihos} \& {Hernquist}}{{Mihos} \&
  {Hernquist}}{1994}]{MH1994}
{Mihos} J.~C.,  {Hernquist} L.,  1994, \apjl, 425, L13

\bibitem[\protect\citeauthoryear{{Mihos} \& {Hernquist}}{{Mihos} \&
  {Hernquist}}{1996}]{MH1996}
{Mihos} J.~C.,  {Hernquist} L.,  1996, \apj, 464, 641


\bibitem[\protect\citeauthoryear{{Moore}, {Ghigna}, {Governato}, {Lake},
  {Quinn}, {Stadel} \& {Tozzi}}{{Moore} et~al.}{1999}]{Moore1999}
{Moore} B.,  {Ghigna} S.,  {Governato} F.,  {Lake} G.,  {Quinn} T.,  {Stadel}
  J.,    {Tozzi} P.,  1999, \apjl, 524, L19

\bibitem[\protect\citeauthoryear{{Naab} \& {Burkert}}{{Naab} \&
  {Burkert}}{2003}]{NB2003}
{Naab} T.,  {Burkert} A.,  2003, \apj, 597, 893

\bibitem[\protect\citeauthoryear{{Navarro} \& {White}}{{Navarro} \&
{White}}{1994}]{NW1994}
{Navarro} J.~F,{White} S.~D.~M., 1994, \mnras, 267 , 401

\bibitem[\protect\citeauthoryear{{Negroponte} \& {White}}{{Negroponte} \&
  {White}}{1983}]{NW1983}
{Negroponte} J.,  {White} S.~D.~M.,  1983, \mnras, 205, 1009

\bibitem[\protect\citeauthoryear{{Patton}, {Pritchet}, {Carlberg}, {Marzke},
  {Yee}, {Hall}, {Lin}, {Morris}, {Sawicki}, {Shepherd} \& {Wirth}}{{Patton}
  et~al.}{2002}]{Patton2002}
{Patton} D.~R.,  {Pritchet} C.~J.,  {Carlberg} R.~G.,  {Marzke} R.~O.,  {Yee}
  H.~K.~C.,  {Hall} P.~B.,  {Lin} H.,  {Morris} S.~L.,  {Sawicki} M.,
  {Shepherd} C.~W.,    {Wirth} G.~D.,  2002, \apj, 565, 208

\bibitem[\protect\citeauthoryear{{Pfrommer}, {Springel}, {En{\ss}lin} \&
  {Jubelgas}}{{Pfrommer} et~al.}{2006}]{Pfrommer2006}
{Pfrommer} C.,  {Springel} V.,  {En{\ss}lin} T.~A.,    {Jubelgas} M.,  2006,
  \mnras, 367, 113

\bibitem[\protect\citeauthoryear{{Sanders}, {Soifer}, {Elias}, {Madore},
  {Matthews}, {Neugebauer} \& {Scoville}}{{Sanders} et~al.}{1988}]{Sanders1988}
{Sanders} D.~B.,  {Soifer} B.~T.,  {Elias} J.~H.,  {Madore} B.~F.,  {Matthews}
  K.,  {Neugebauer} G.,    {Scoville} N.~Z.,  1988, \apj, 325, 74

\bibitem[\protect\citeauthoryear{{Somerville} \& {Primack}}{{Somerville} \&
  {Primack}}{1999}]{SP1999}
{Somerville} R.~S.,  {Primack} J.~R.,  1999, \mnras, 310, 1087

\bibitem[\protect\citeauthoryear{{Somerville}, {Primack} \&
  {Faber}}{{Somerville} et~al.}{2001}]{Somerville2001}
{Somerville} R.~S.,  {Primack} J.~R.,    {Faber} S.~M.,  2001, \mnras, 320, 504

\bibitem[\protect\citeauthoryear{{Spergel}, {Verde}, {Peiris}, {Komatsu},
  {Nolta}, {Bennett}, {Halpern}, {Hinshaw}, {Jarosik}, {Kogut}, {Limon},
  {Meyer}, {Page}, {Tucker}, {Weiland}, {Wollack} \& {Wright}}{{Spergel}
  et~al.}{2003}]{Spergel2003}
{Spergel} D.~N.,  {Verde} L.,  {Peiris} H.~V.,  {Komatsu} E.,  {Nolta} M.~R.,
  {Bennett} C.~L.,  {Halpern} M.,  {Hinshaw} G.,  {Jarosik} N.,  {Kogut} A.,
  {Limon} M.,  {Meyer} S.~S.,  {Page} L.,  {Tucker} G.~S.,  {Weiland} J.~L.,
  {Wollack} E.,    {Wright} E.~L.,  2003, \apjs, 148, 175


\bibitem[\protect\citeauthoryear{{Springel}, {Di Matteo} \&
  {Hernquist}}{{Springel} et~al.}{2005a}]{volker2005b}
{Springel} V.,  {Di Matteo} T.,    {Hernquist} L.,  2005a, \apjl, 620, L79

\bibitem[\protect\citeauthoryear{{Springel} \& {Hernquist}}{{Springel} \&
  {Hernquist}}{2003}]{SH2003}
{Springel} V.,  {Hernquist} L.,  2003, \mnras, 339, 289

\bibitem[\protect\citeauthoryear{{Springel}, {White}, {Jenkins}, {Frenk},
  {Yoshida}, {Gao}, {Navarro}, {Thacker}, {Croton}, {Helly}, {Peacock}, {Cole},
  {Thomas}, {Couchman}, {Evrard}, {Colberg} \& {Pearce}}{{Springel}
  et~al.}{2005b}]{Nature2005}
{Springel} V.,  {White} S.~D.~M.,  {Jenkins} A.,  {Frenk} C.~S.,  {Yoshida} N.,
   {Gao} L.,  {Navarro} J.,  {Thacker} R.,  {Croton} D.,  {Helly} J.,
  {Peacock} J.~A.,  {Cole} S.,  {Thomas} P.,  {Couchman} H.,  {Evrard} A.,
  {Colberg} J.,    {Pearce} F.,  2005b, \nat, 435, 629

\bibitem[\protect\citeauthoryear{{Springel}, {White}, {Tormen} \&
  {Kauffmann}}{{Springel} et~al.}{2001}]{Springel2001b}
{Springel} V.,  {White} S.~D.~M.,  {Tormen} G.,    {Kauffmann} G.,  2001,
  \mnras, 328, 726


\bibitem[\protect\citeauthoryear{{Toomre}}{{Toomre}}{1976}]{1976T}
{Toomre} A.,  1976, in Bulletin of the American Astronomical Society Vol.~8 of
  Bulletin of the American Astronomical Society, {INVITED PAPER - Interacting
  Galaxies}.
pp 354--+

\bibitem[\protect\citeauthoryear{{van Dokkum}}{{van Dokkum}}{2005}]{Dokkum2005}
{van Dokkum} P.~G.,  2005, \aj, 130, 2647

\bibitem[\protect\citeauthoryear{{Velazquez} \& {White}}{{Velazquez} \&
  {White}}{1999}]{VW1999}
{Velazquez} H.,  {White} S.~D.~M.,  1999, \mnras, 304, 254

\bibitem[\protect\citeauthoryear{{White}}{{White}}{1978}]{1978W}
{White} S.~D.~M.,  1978, \mnras, 184, 185

\bibitem[\protect\citeauthoryear{{White} \& {Rees}}{{White} \&
  {Rees}}{1978}]{WR1978}
{White} S.~D.~M.,  {Rees} M.~J.,  1978, \mnras, 183, 341

\end{thebibliography}

\end {document}